\documentclass{LT23auth}
\usepackage{graphicx}

\begin{document}

\begin{frontmatter}

\title{Peculiar roles of spins in the thermal conductivity
of pure and doped La$_{2}$CuO$_4$: Comparison with CuGeO$_3$}

\author[address1]{Yoichi Ando \thanksref{thank1}},
\author[address1]{X. F. Sun},
\author[address1]{J. Takeya},
\author[address1]{Seiki Komiya}

\address[address1]{Central Research Institute of Electric Power Industry,
Komae, Tokyo 201-8511, Japan}

\thanks[thank1]{Corresponding author. E-mail: ando@criepi.denken.or.jp}

\begin{abstract}

In-plane and out-of-plane thermal conductivities ($\kappa_{ab}$ and
$\kappa_{c}$) are measured on single crystals of pure, 1\%-hole-doped,
and 1\%-Zn-doped La$_{2}$CuO$_4$. The roles of magnons and the spin
stripes in the heat transport in these samples are discussed. Comparison
with the heat transport in CuGeO$_3$, which shows similar $\kappa(T)$
behavior as that of La$_{2}$CuO$_4$, gives us a lesson of how the heat
transport can probe the difference in the spin ground state.

\end{abstract}

%
%
\end{frontmatter}

Thermal conductivity is a basic transport property that usually bears
information on charge carriers and phonons, as well as their scattering
processes. In materials where spins are playing a major role, the
behavior of the thermal conductivity is largely governed by the spins,
because the spin excitations can carry heat current themselves and also
scatter electrons and phonons. Although the heat transport due to spin
excitations in magnetic materials has been known for a long time
\cite{Dixon}, recent studies of the heat transport in
strongly-correlated low-dimensional cuprate systems (such as
high-$T_\mathrm{c}$ cuprates \cite{Nakamura}, spin-Peierls material
CuGeO$_3$ \cite{CuGeO,Takeya}, {\it etc.}) have
found the spin-related heat transport to be useful for extracting
information on the peculiar spin systems in these compounds. In this
paper, we present some new data on the thermal conductivity $\kappa$ of
lightly Sr- or Zn-doped La$_{2}$CuO$_4$ (LCO), where we find unusual
difference between Sr and Zn doping. Comparison of the data of
lightly-doped LCO with those of CuGeO$_3$ demonstrates how the behavior
of $\kappa(T)$ reflects the difference in the spin ground states.

The single crystals of LCO, La$_{2-x}$Sr$_x$CuO$_4$ (LSCO), and
La$_2$Cu$_{1-y}$Zn$_y$O$_4$ (LCZO) are grown by the traveling-solvent
floating-zone technique \cite{mobility}. The thermal conductivity is
measured with a steady-state technique below 150 K, and with a modified
steady-state technique, which minimizes the radiation loss, above 150 K.
Details of the CuGeO$_3$ experiments are described in Refs.
\cite{CuGeO} and \cite{Takeya}.

\begin{figure}[btp]
\begin{center}\leavevmode
\includegraphics[width=0.7\linewidth]{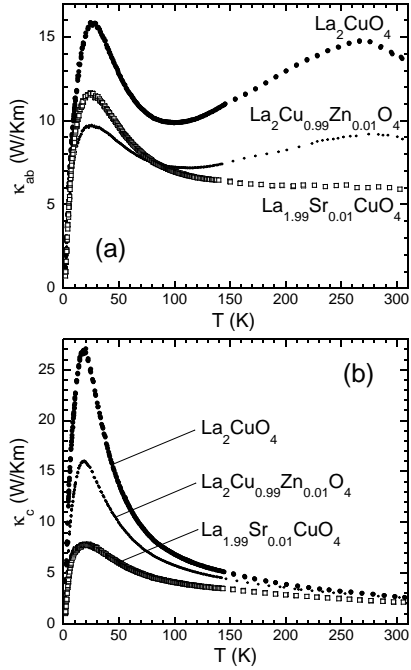}
\caption{(a) $\kappa_{ab}(T)$ and (b) $\kappa_c(T)$ data of LCO, 
LSCO ($x$=0.01), and LCZO ($y$=0.01).}
\label{Fig1}\end{center}\end{figure}

Figure 1 shows $\kappa_{ab}(T)$ and $\kappa_c(T)$ for LCO, LSCO
($x$=0.01), and LCZO ($y$=0.01); here, we concentrate on comparing the
effect of hole doping and Zn doping of the same amount, 1\%. The LCO
sample has the N\'{e}el temperature $T_\mathrm{N}$ of 307 K. It has been
discussed \cite{Nakamura} for LCO that the high-temperature peak in
$\kappa_{ab}(T)$ is due to magnons, while the low-temperature peak is
due to phonons. In our doped samples, $T_\mathrm{N}$ is still relatively
high for both LSCO with $x$=0.01 ($T_\mathrm{N}$=242 K) and LCZO with
$y$=0.01 ($T_\mathrm{N}$=295 K); nevertheless, the high-temperature peak
already disappears in LSCO, while it clearly remains in LCZO. This
observation supports the picture that magnons cause the high-temprature
peak and suggests that Sr doping is more detrimental to the magnon 
heat transport than Zn doping. On the other
hand, the low-temperature phonon peak in $\kappa_{ab}(T)$ is more
strongly suppressed in LCZO than in LSCO; this is not what is expected
from simple defect scattering, because the atomic mass difference
between Cu and Zn is much smaller than that between La and Sr. Most
likely, the enhanced damping of the phonon peak in LCZO is due to the
magnon scattering of phonons.

When one looks at the $\kappa_c(T)$ behaviors in Fig. 1(b), one notices
that the difference between Sr- and Zn-doping is more pronounced in this
direction, and the trend is opposite to that in $\kappa_{ab}(T)$;
namely, along the $c$-axis the Sr doping causes much stronger damping of
the phonon peak. Although the exact mechanism of this strong damping is
to be scrutinized by future research, we tentatively ascribe this effect
to the spin stripes observed by neutrons \cite{Matsuda} in this lightly
Sr-doped region: Since there is a strong spin-lattice coupling in LSCO
\cite{Lavrov}, formation of the spin stripes is expected to induce local
lattice distortions; such lattice distortions do not scatter phonons if
they are periodic (in fact, we do not see strong scattering of phonons
in the in-plane direction, for which the stripes are well ordered), but
the stripes are known to be very disordered along the $c$-axis
\cite{Matsuda}. Thus, we have a good reason to expect that the anomalous
damping of the phonon peak in $\kappa_c$(T) with only 1\% of hole doping
is due to the existence of peculiar spin texturing and strong
spin-lattice coupling.

\begin{figure}[btp]
\begin{center}\leavevmode
\includegraphics[width=0.75\linewidth]{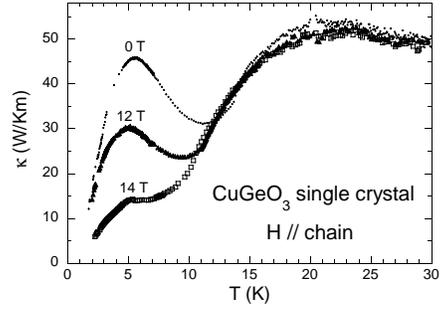}
\caption{$\kappa(T)$ data of CuGeO$_3$, taken from Ref. \cite{CuGeO}, 
along the chain direction. The magnetic field is applied along the chains.}
\label{Fig2}\end{center}\end{figure}

Comparison of the data in Fig. 1(a) to the $\kappa(T)$ behavior of
CuGeO$_3$ \cite{CuGeO} measured along the spin chain direction (Fig. 2)
is quite illuminating. The 0-T data in Fig. 2 look quite similar to the
$\kappa_{ab}(T)$ data of LCO, which is natural because in both systems
the high-temperature peak is caused by the heat conduction due to spin
excitations and the low-temperature peak is due to phonons. The spin heat 
transport (high-temperature peak) is quickly suppressed upon doping 
Mg to CuGeO$_3$ \cite{Takeya}, as is the case with doped LCO. The
difference between the two systems becomes evident when one compares the
magnetic-field dependences. In CuGeO$_3$, as is shown in Fig. 2,
application of magnetic field along the chain direction causes a damping
of the phonon peak, and the 14-T field, which closes the spin gap of the
spin-Peierls state, completely suppresses the phonon peak; we discussed
\cite{CuGeO} that this is because more and more spin excitations (which
scatter phonons) are allowed as the spin gap is suppressed with the
magnetic field. On the other hand, for LCO, we observed that the
$\kappa(T)$ behavior is essentially unchanged with magnetic field;
this is understandable, because the characteristic magnetic energy in
LCO is given by the exchange interaction $J$, which is order of 2000 K.
Therefore, the behavior of the thermal conductivity clearly reflects the
different nature of the spin ground state in the two systems.

%
%
\begin{ack}
The experiments on CuGeO$_3$ were done in collaboration with 
A. Kapitulnik, I. Tanaka, R. S. Feigelson, I. Tsukada, T. Masuda,
and K. Uchinokura.
\end{ack}

%
%


\begin{thebibliography}{9}

\bibitem{Dixon} 
G. S. Dixon, D. P. Landau, Phys. Rev. B {\bf 13} (1976) 3121.

\bibitem{Nakamura} 
Y. Nakamura {\it et al.}, Physica C {\bf 185-189} (1991) 1409. 

\bibitem{CuGeO} 
Y. Ando {\it et al.}, Phys. Rev. B {\bf 58} (1998) R2913. 

\bibitem{Takeya}
J. Takeya {\it et al.}, Phys. Rev. B {\bf 61} (2000) 14700.

\bibitem{mobility}
Y. Ando {\it et al.}, Phys. Rev. Lett. {\bf 87} (2001) 017001. 

\bibitem{Matsuda} 
M. Matsuda {\it et al.}, Phys. Rev. B {\bf 65} (2002) 134515. 

\bibitem{Lavrov}
A. N . Lavrov, S. Komiya, Y. Ando, Nature {\bf 418} (2002) 385.

\end{thebibliography}
\end{document}